\documentclass[preprint,aps,prd,nofootinbib,superscriptaddress]{revtex4-1} 
\pdfoutput=1

\usepackage{graphicx} 
\usepackage{subfigure}
\usepackage{amsmath}

\begin{document}
\title{The Effective Hooperon}
\author{Alexandre Alves}
\affiliation{Departamento de Ci\^encias Exatas e da Terra, Universidade Federal de S\~ao Paulo, Diadema-SP, 09972-270, Brasil}
\author{Stefano Profumo}
\affiliation{Santa Cruz Institute for Particle Physics and Department of Physics, University of California Santa Cruz, Santa Cruz CA, USA}
\author{Farinaldo S. Queiroz}
\affiliation{Santa Cruz Institute for Particle Physics and Department of Physics, University of California Santa Cruz, Santa Cruz CA, USA}
\author{William Shepherd}
\affiliation{Santa Cruz Institute for Particle Physics and Department of Physics, University of California Santa Cruz, Santa Cruz CA, USA}

\date{\today}

\begin{abstract}
\noindent We explore the possibility of explaining a gamma-ray excess in the Galactic Center, originally pointed out by Hooper, collaborators, and other groups, in an effective field theory framework. We assume that dark matter  annihilation is mediated by particles heavy enough to be integrated out, and that such particles couple to all quark families. We calculate the effective coupling required to explain the annihilation signal in the Galactic Center, and compare with bounds from direct detection, collider searches, and the requirement that the dark matter particle make up the appropriate fraction of the universal energy budget. We find that only a very small set of operators can explain the gamma-ray excess while being consistent with other constraints. Specifically, for scalar dark matter the viable options are one scalar-type coupling to quarks and one interaction with gluons, while for fermionic (Dirac) dark matter the viable options are two scalar-type dimension-7 operators or a dimension-6 vector-type operator. In all cases, future searches with the Large Hadron Collider should probe the relevant operators' effective energy scale, while all viable interactions should escape direct detection experiments. 
\end{abstract}

\maketitle

\section{Introduction}
\label{sec:intro}

The gamma-ray flux from the Galactic Center region contains an excess, at energies in the 1-10 GeV range, over standard choices for the astrophysical background, as noted in 2009 by Goodenough and Hooper \cite{Goodenough:2009gk}, reiterated in 2010 by the same authors \cite{Hooper:2010mq}, then in 2011 by Hooper and Linden \cite{Hooper:2011ti}, and recently again by Hooper and collaborators \cite{newhooperon} and by other groups \cite{Abazajian:2012pn, Macias:2013vya, Abazajian:2014fta}. The Fermi-LAT Collaboration has also searched the Galactic Center \cite{fermilatgc}, reaching less striking conclusions. The excess has been tentatively associated with the pair-annihilation of a weakly-interacting massive particle (WIMP) in the inner Galaxy. For natural reasons, we indicate the tentative dark matter (DM) particle whose annihilation would produce said excess as the ``Hooperon''. If, indeed, astrophysical backgrounds cannot reproduce the observed excess, such a discovery would be an exciting breakthrough in the ongoing attempt to understand the nature of DM in the Universe.  Pinpointing the implications of this scenario for particle physics model building might be perhaps premature, but could eventually become an exercise of the utmost importance to shed light on physics beyond the Standard Model (SM).

Generally, to learn something about the implications of a DM model one requires a complete understanding of its interactions with SM particles. In some cases, however, it is possible to model such interactions using a relatively small number of possible operators, so long as the force-mediating particles are much heavier than the DM particles, and can thus be effectively integrated out. The effective field theory (EFT) framework necessary to perform such an analysis has been thoroughly explored in the literature (see e.g. \cite{everyone, Goodman:2010ku, Profumo:2013hqa}). Here, we apply EFT  techniques to model a DM ``Hooperon'' particle that could produce the observed Galactic Center excess while being consistent with collider searches, direct DM searches, and producing an acceptable universal DM density in the early Universe. Indirect detection bounds are also relevant and are on the verge of probing the excess discussed here \cite{dwarfsbounds,GCbounds}.

The present study is organized as follows: In the next section, we succinctly review the EFT framework; In Sec.~\ref{sec:bounds} we summarize  current constraints on the relevant EFT operators from direct detection, relic density considerations, and collider searches; In Sec.~\ref{sec:signal} we calculate the required interaction strengths to reproduce annihilation cross sections consistent with the Galactic Center excess, and compare those values with current bounds; Finally, we conclude in Sec.~\ref{sec:conc}.

\section{Model Independent Dark Matter Interactions}
\label{sec:effth}

Under the assumption that the particles mediating interactions of DM with SM fields are very heavy, those fields can be naturally integrated out of the theory, leaving as their only low-energy counterpart a set of contact operators which parametrize the interactions of DM with the SM. If we consider only those operators which are least suppressed (of dimension less than or equal to 7) by the heavy mass scale we get a manageable number of candidate interaction terms. This interaction basis very naturally interfaces with results from direct detection experiments, where any new physics even marginally heavier than the DM is too heavy to see directly due to the small kinetic energies involved in DM scattering. Extrapolating the same operators to the energy scale of collisions at the LHC is somewhat less solid ground, but it still gives models which can be successfully searched for. If we find that  LHC bounds rule out a specific operator that otherwise would explain the Galactic Center excess then we can conclude that, in order for that particular explanation to be valid, the heavy-mediator approximation must break down \cite{lightmed}. 

We point out that the relevant scales in our setup are lower than those where it is certain that the EFT approximation is valid for collider searches. We believe, however, that these models provide an important benchmark which can be aimed for in DM searches, and include many of the interesting effects that can manifest themselves in DM physics, in particular chirality and p-wave 
suppressions of various processes. It is very important to note that LHC bounds do not
necessarily weaken as one leaves the EFT limit, they can also be notably stronger. Thus, having a benchmark to search for is quite valuable. Of course, given a more complete model of DM physics, it would be better to design LHC searches to suit its particular features, but the surfeit of such models makes a more simple case worth considering. Also, suppression scales of order 100 GeV can easily be due to new physics at strikingly higher scales which is coupled somewhat strongly and therefore will not manifest itself as more than a contact operator in low-energy searches like direct and indirect detection. The aim of this work is precisely to provide a simple benchmark for the physics which could give a DM explanation of the Galactic Center excess and refer to the corresponding direct detection and collider constraints. Our general conclusions based on this EFT approach agree well with the simplified model attempt in Ref.\cite{hooperon}.

That being said, the possible interactions we consider are listed in table \ref{tab:ops}. Note that motivated by the observed spectral features of the Galactic Center gamma-ray excess, we consider only couplings to hadronic particles, and assume there is no coupling of DM to leptons or directly to electroweak bosons. While it is possible to explain the Galactic Center excess with annihilations into $\tau$ leptons, it is not as good a fit as the hadronic channels because annihilations into $\tau$ leptons yields a softer gamma-ray spectrum in the energy of interest \cite{Hooper:2011ti}, and thus we do not consider it further. We point out that when inverse Compton scattering processes off galactic photon background are included leptonic annihilations provide a good fit  \cite{Lacroix:2014eea}. Although the annihilation cross section is this seems to be in tension with AMS02 exclusion limits \cite{Bergstrom:2013jra}. Anyway, we assume hereafter that only hadronic annihilations take place and draw our conclusions under that assumption.

 We adopt the operator naming convention of Ref.~\cite{Goodman:2010ku}. This list of operators is a complete basis of all interactions of DM with hadronic matter within the heavy-mediator limit. Leaving that limit has been explored in detail as well, but involves more assumptions about the DM model \cite{lightmed}. Each operator is preceded by an assumed Wilson coefficient. The coefficients for operators D1-4 and C1 and 2 scale with the quark mass in recognition of their violation of SM chiral symmetries; this can be considered as a Higgs field which has been set to its vacuum expectation value. Operators D5-8 and C3 and 4 do not violate chirality, and so do not carry any mass scaling.  The tensor operators, D9 and D10 as written, connect to the angular momentum of the nucleon due to quarks. While they do violate SM chirality, we do not scale them by quark masses to make contact with the hadronic spin variables. The Wilson coefficients for D11-14 and C5 and 6 include a factor of the strong coupling constant to account for the expectation that they are loop-induced. All of these Wilson coefficients are the usual choices within the literature on effective theory treatments of DM \cite{everyone, Goodman:2010ku}

\begin{table}
\subtable[\ Operators for Dirac fermion DM]{\begin{tabular}{|l|c|c|c|}
\hline
Name&Operator&Dimension&SI/SD\\
\hline
D1&$\frac{m_q}{\Lambda^3}\bar\chi\chi\bar qq$&7&SI\\
D2&$\frac{im_q}{\Lambda^3}\bar\chi\gamma^5\chi\bar qq$&7&N/A\\
D3&$\frac{im_q}{\Lambda^3}\bar\chi\chi\bar q\gamma^5q$&7&N/A\\
D4&$\frac{m_q}{\Lambda^3}\bar\chi\gamma^5\chi\bar q\gamma^5q$&7&N/A\\
D5&$\frac{1}{\Lambda^2}\bar\chi\gamma^\mu\chi\bar q\gamma_\mu q$&6&SI\\
D6&$\frac{1}{\Lambda^2}\bar\chi\gamma^\mu\gamma^5\chi\bar q\gamma_\mu q$&6&N/A\\
D7&$\frac{1}{\Lambda^2}\bar\chi\gamma^\mu\chi\bar q\gamma_\mu\gamma^5q$&6&N/A\\
D8&$\frac{1}{\Lambda^2}\bar\chi\gamma^\mu\gamma^5\chi\bar q\gamma_\mu\gamma^5q$&6&SD\\
D9&$\frac{1}{\Lambda^2}\bar\chi\sigma^{\mu\nu}\chi\bar q\sigma_{\mu\nu}q$&6&SD\\
D10&$\frac{i}{\Lambda^2}\bar\chi\sigma^{\mu\nu}\gamma^5\chi\bar q\sigma_{\mu\nu}q$&6&N/A\\
D11&$\frac{\alpha_s}{\Lambda^3}\bar\chi\chi G^{\mu\nu}G_{\mu\nu}$&7&SI\\
D12&$\frac{\alpha_s}{\Lambda^3}\bar\chi\gamma^5\chi G^{\mu\nu}G_{\mu\nu}$&7&N/A\\
D13&$\frac{\alpha_s}{\Lambda^3}\bar\chi\chi G^{\mu\nu}\tilde{G}_{\mu\nu}$&7&N/A\\
D14&$\frac{\alpha_s}{\Lambda^3}\bar\chi\gamma^5\chi G^{\mu\nu}\tilde{G}_{\mu\nu}$&7&N/A\\
\hline
\end{tabular}}
\qquad
\subtable[\ Operators for Complex scalar DM]{\begin{tabular}{|l|c|c|c|}
\hline
Name&Operator&Dimension&SI/SD\\
\hline
C1&$\frac{m_q}{\Lambda^2}\phi^\dagger\phi\bar qq$&6&SI\\
C2&$\frac{m_q}{\Lambda^2}\phi^\dagger\phi\bar q\gamma^5q$&6&N/A\\
C3&$\frac{1}{\Lambda^2}\phi^\dagger\overleftrightarrow{\partial}_\mu\phi\bar q\gamma^\mu q$&6&SI\\
C4&$\frac{1}{\Lambda^2}\phi^\dagger\overleftrightarrow{\partial}_\mu\phi\bar q\gamma^\mu\gamma^5 q$&6&N/A\\
C5&$\frac{\alpha_s}{\Lambda^3}\phi^\dagger\phi G^{\mu\nu}G_{\mu\nu}$&6&SI\\
C6&$\frac{\alpha_s}{\Lambda^3}\phi^\dagger\phi G^{\mu\nu}\tilde{G}_{\mu\nu}$&6&N/A\\
\hline
\end{tabular}}
\caption{\label{tab:ops}Lowest-dimensional operators which couple singlet DM candidates to hadronic matter. The  column to the right indicates whether the primary direct detection signal due to that operator is spin-independent (SI), spin-dependent (SD), or strongly suppressed (N/A).}
\end{table}

\section{Constraints on Effective DM Models}
\label{sec:bounds}

The most stringent and uncontroversial probes of the interactions mediated by the operators listed in the previous section are direct detection experiments. In particular, spin-independent direct detection searches give strong bounds on operators which can mediate such interactions, and in fact completely rule out the parameter space of interest for all the operators which lead to unsuppressed scattering of that type. The bounds from direct detection on all of the relevant operators require suppression scales at least in the multi-TeV range, far beyond the region of interest to explain the gamma-ray excess \cite{Goodman:2010ku}. We therefore drop these operators as potential explanations of the Galactic Center gamma ray excess, since they are far too strongly constrained to contribute in any meaningful way. Spin-dependent direct detection bounds the operators D8 and D9, but we will find that other considerations disfavor these interactions as possible explanations of the Galactic Center signal.

The main other class of constraints on the effective interactions we consider here comes from collider searches. Generically, at colliders one looks for a SM particle radiated off of the initial state quarks, which then annihilate through the given operator into a DM pair which escapes the detector, and whose presence can only be inferred from the missing transverse momentum in a given event. Searches using many different SM final states have been performed \cite{Aaltonen:2012jb,CMS-PAS-EXO-12-048,Aad:2013oja}. We have selected here the most stringent constraint on each operator. In some cases, stronger bounds are available if the relative sign between the couplings to up- and down-type quarks is allowed to change. This is notably the case for searches for $W$ bosons and missing energy. In these cases we present multiple curves, one for same-sign and one for opposite-sign couplings.

Collider bounds are largely insensitive to the presence or absence of $\gamma^5$ factors in the bilinears of quarks or DM particles. This is because all particles in the collider have large velocities, and thus are in definite helicity states, so that there can be no interference between left- and right-handed particles, unlike the case of small velocities of relevance for both annihilations and direct detection scattering events. Thus, we will present a single collider curve for each class of operators. These bounds are included in figures \ref{fig:dirac} and \ref{fig:scalar}. While the loop-induced mixing of operators D1-4 into D11-14 can give a strong enhancement to the collider bounds on the former \cite{Haisch:2012kf}, we conservatively present only the tree-level result here.

It is worth keeping in mind that the limit in which the EFT approach to DM interactions is valid is violated first in collider searches, where large momenta can be imparted to the produced DM pair. Thus, it is entirely reasonable to believe that, while the low-momentum interactions of DM with the SM could be described by contact operators, the collider search bounds could be significantly altered. Such cases have been researched in detail by many authors \cite{lightmed}, who find that the collider constraints can be appreciably stronger or weaker in different regions of the light mediator parameter space.

Additionally, we also show in figures \ref{fig:dirac} and \ref{fig:scalar} the line in parameter space which corresponds to a  relic DM density compatible with observation, under the assumption that no other operator besides the one under consideration is appreciably contributing to the annihilation rate. Finally, we present dotted direct detection exclusion curves in each of the figures. For the operators D3, D4, D7, C2 and C6 those bounds are irrelevant for the region of interest and thus ignored in agreement with \cite{DelNobile:2013sia}.

The usual interpretation of such a curve in more complete models is that annihilation must be at least efficient enough to keep from predicting too large a relic density, because the model would then be excluded by  precise measurements of the energy density of matter in the Universe \cite{Ade:2013zuv}.
However, in the present case the situation is slightly different: It is always possible to enhance the annihilation cross section by turning on another operator, and therefore to dilute the DM relic density to be the correct value, without spoiling the results we find (in other words: invoking an operator that does not contribute to the low-velocity annihilation processes but that does contribute to annihilation in the early Universe). For instance, if we consider the operator D2 as an interesting possible explanation of the Galactic Center excess, but its s-wave annihilations are not sufficient to dilute the DM relic density sufficiently, we can consider an admixture of D2 and D3, where the purely p-wave annihilations given by D3 help increase the effective annihilation rate in the early Universe without changing the rate we observe today. This could be simply achieved in a one-particle UV-completion of the model by having a spin-0 mediator which couples as a pseduoscalar to DM but with more involved chiral couplings to the quarks. Note that this would require a less-suppressed D3 interaction than that for D2, and would run afoul of the LHC bounds on D3 in the heavy-mediator limit, but collider-accesibility of the mediating particle may reduce the LHC sensitivity to these interactions. Likewise the operators D1, D6, D11, D13, C3, and C4 also give purely p-wave suppressed annihilations, and could be used to reduce the relic abundance of DM without affecting its current annihilation signals. Additionally, other annihilation channels to non-hadronic particles could easily be envisioned, and would allow the relic density to be satisfied while retaining the same signal strength in hadronic final states from the Galactic Center. 

So long as we remain within a standard thermal history of the Universe, nothing can instead be done to increase the DM relic density if our preferred model predicts too little. Of course, non-standard cosmological histories can either increase or decrease the DM density compared to the thermal relic expectation. While we might attempt to resolve this issue by appealing to an additional component of the DM energy density not responsible for the gamma ray excess, it is actually impossible to match the excess at all with a model which predicts a too-small relic density. If we were to imagine that the annihilation cross section is simply larger by the needed factor to overcome the reduced abundance, we would reduce the abundance further (the annihilation rate scales with the thermally averaged pair-annihilation cross section times velocity in the same way as the density: $\rho_{\rm DM}\sim\langle\sigma v\rangle^{-1}$ and $\Gamma_{\rm ann}\sim\rho_{\rm DM}^2\langle\sigma v\rangle\sim\langle\sigma v\rangle^{-1}$). Iteratively trying to correct for the decreasing abundance will force the model to stronger and stronger interactions, eventually predicting effectively zero abundance of the would-be DM candidate. Thus, we will only be interested in points in parameter space where the interaction is too weak to give the correct relic density or ideal, and we consider interactions stronger than those needed to give the relic density to be ruled out as possible explanations of this signal.

\section{Explaining the Galactic Center Excess}
\label{sec:signal}

\begin{figure}
\subfigure[\label{fig:D2-4}\ Scalar-Type Operators]{\includegraphics[width=0.45\textwidth]{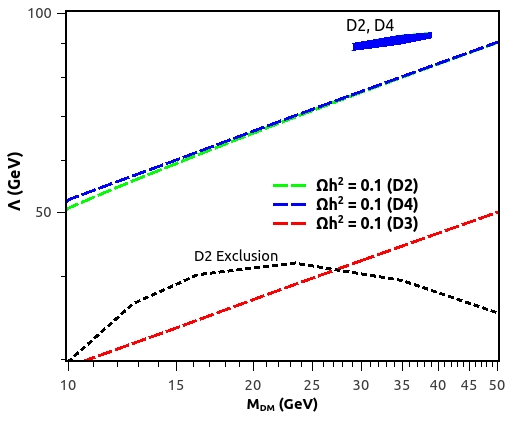}}
\subfigure[\label{fig:D6-8}\ Vector-Type Operators]{\includegraphics[width=0.45\textwidth]{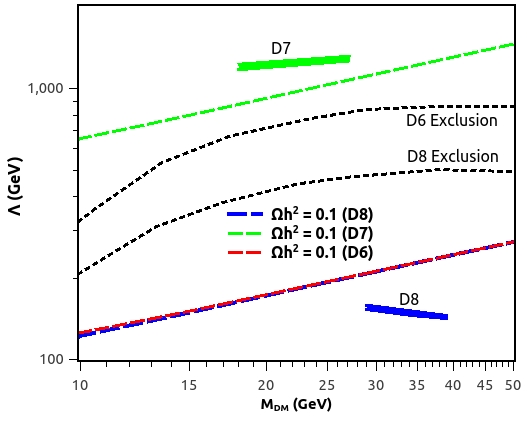}}
\subfigure[\label{fig:D9-10}\ Tensor-Type Operators]{\includegraphics[width=0.45\textwidth]{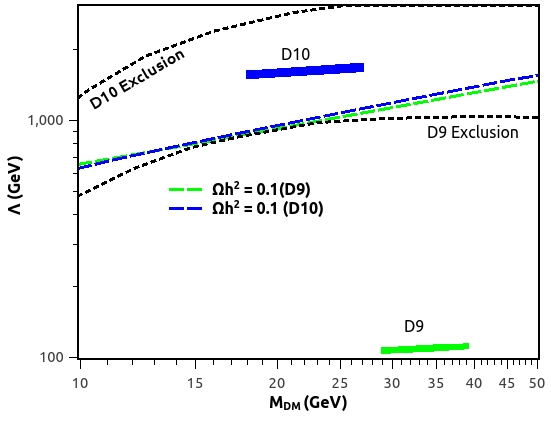}}
\subfigure[\label{fig:D12-14}\ Gluonic Operators]{\includegraphics[width=0.45\textwidth]{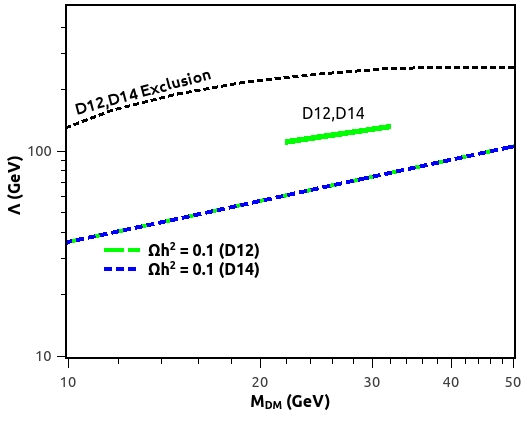}}
\caption{\label{fig:dirac} Bounds on effective theory parameters and favored regions to explain the Galactic Center gamma-ray excess from Dirac DM. LHC bounds are shown as black lines (the excluded regions are below the lines); dashed  lines indicate the case of coupling choices made to maximize constructive interference, while solid lines indicate generic bounds which apply equally to all of the relevant operators. Colored dashed lines indicate the parameter space where the effective operators produce the correct thermal relic density of DM, while the regions corresponding to good fits to the gamma-ray excess are shown as filled areas with the same color. The right abundance lines yield $\Omega h^2=0.12$. The abundances for the favored regions are: $0.13 \leq \Omega h^2 (D_2,D_4) \leq 0.15$, $0.16 \leq \Omega h^2 (D_7) \leq 0.18$, $0.05 \leq \Omega h^2 (D_8) \leq 0.07$,$1\times 10^{-5} \leq \Omega h^2 (D_9) \leq 2 \times 10^{-5}$,$0.26 \leq \Omega h^2 (D_{10}) \leq 0.285$,$0.45 \leq \Omega h^2 (D_{12},D_{14}) \leq 0.49$.}
\end{figure}

\begin{figure}
\subfigure[\label{fig:C2}Scalar-Type Operator]{\includegraphics[width=0.45\textwidth]{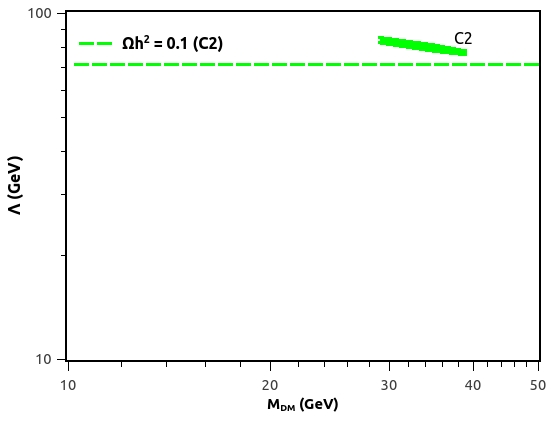}}
\subfigure[\label{fig:C6}Gluon-Coupling Operator]{\includegraphics[width=0.45\textwidth]{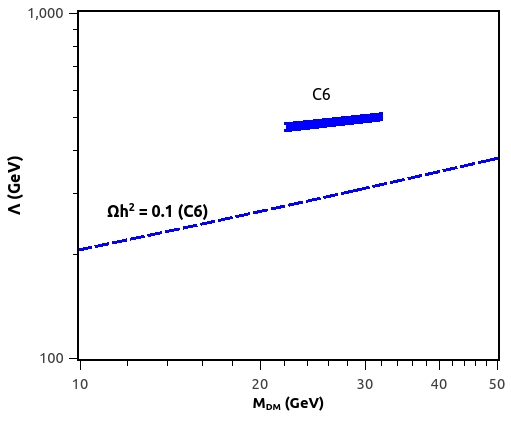}}
\caption{\label{fig:scalar}Bounds on effective theory parameters and favored regions to explain the Galactic Center gamma-ray excess from complex scalar DM. LHC bounds are shown as solid black lines, the correct relic density corresponds to the dashed lines, and the thick regions provide the correct annihilation cross section to explain the Galactic Center signal. The right abundance lines yield $\Omega h^2=0.12$. The abundances for the favored regions are: $0.128 \leq \Omega h^2 (C_2) \leq 0.14$,$0.434 \leq \Omega h^2 (C_6) \leq 0.445$}
\end{figure}

The coupling structures that we consider with these effective operators have been characterized as either ``democratic'' or ``mass-coupled'' in \cite{newhooperon}, and the annihilation cross section into quarks preferred as a function of mass is given therein for each of these scenarios. We calculate the suppression scale which gives the required annihilation cross section for each operator under consideration, and show the resulting regions in figures \ref{fig:dirac} and \ref{fig:scalar}. Na\"{\i}vely we expect that operators D2-4 and C2 give rise to the ``mass-coupled'' pattern of final state particles and D6-10 and C4 to the ``democratic'' scenario, but it is worth noting that annihilations through D8 and D9 are actually helicity-suppressed, and thus also give rise to a ``mass-coupled'' pattern of branching fractions as well, despite having no quark mass in their actual couplings. We are also considering operators which allow DM to annihilate to gluon pairs rather than quark pairs. Gluons are known to give gamma-ray spectra which are nearly identical to those from charm quarks \cite{Cirelli:2010xx}, so we extract the cross-sections and masses which are preferred for charm final states from \cite{newhooperon} and apply those criteria to the operators D12-14 and C6.

The results for scalar-type couplings of fermionic DM are shown in figure \ref{fig:D2-4}. The LHC bound stemming from a hadronic W or Z boson and missing energy search at ATLAS \cite{Aad:2013oja} excluded $\Lambda < 60$~GeV. However is debatable the applicability of this limit.    
The dashed lines indicate the points with the correct thermal relic density. As explained above, regions below these dashed lines are ruled out (under the assumption of absence of non-thermal production of DM). We have not plotted results for D1, which is strongly constrained by direct detection, but we have plotted the results for all other operators of this class. We also show the direct detection limit on the operator D2 with a dotted black line. For D3,D4 the bounds are weak and thus not presented in the figure.  

The operators D2 and D4 give identical favored regions to explain the Galactic Center anomaly, and those regions are not constrained by LHC searches or by the requirement that the relic density possibly be satisfied. Note, however, that one-loop effects, if taken in to account, will cause the LHC bounds to be more similar to those on gluonic operators  \cite{Haisch:2012kf}, which could potentially rule out the  regions of interest. The direct detection ones are irrelevant as one can clearly see. in summary, D2-D4 are viable operators.

The operator D3 requires extremely low suppression scales (at a scale of about 6 to 7 GeV) to explain the Galactic Center signal because annihilations through this operator are p-wave suppressed, and thus falls far below the region plotted here. This places the favored region in very strong tension with LHC limits and with the relic density requirement. 

Vector-type couplings of fermionic DM are presented in figure \ref{fig:D6-8}. There are two LHC bounds that could be drawn on this space, depending on the relative sign between the coupling to up-type and down-type quarks. If that sign is negative then $\Lambda < 1.9$~TeV is excluded out. 
It is made stronger because the opposite sign leads to constructive interference in the process of emitting a W boson and DM pair, and this bound is from the hadronic W or Z search at ATLAS \cite{Aad:2013oja}. The second bound, 
from a monojet search at CMS \cite{CMS-PAS-EXO-12-048}, and is insensitive to the relative sign in couplings to different types of quarks and rules out $\Lambda > 900$~GeV. The direct detection limits rule out the D6 and D8 operators, but the D7 operator remains a plausible solution.

We see, again, that p-wave annihilating operators, notably D6 which would require a suppression scale of about 30 GeV, are unable to explain the gamma-ray excess. Also we note that D8, while not fully p-wave suppressed, has a helicity-suppressed s-wave annihilation, and thus also underpredicts the relic abundance in the region where the present-day annihilation cross section matches the observed excess. D7 gives an unsuppressed annihilation rate in the late Universe, and thus is not in tension with the requirements of a correct relic density. LHC bounds also disfavor D6 and D8 as explanations of the gamma ray excess in the absence of a light mediating particle. D7 is either allowed or excluded by LHC searches, depending on whether the DM interacts with equal or opposite couplings to up- and down-type quarks. If the new physics is isospin-blind, then D7 is permitted by all current bounds as an explanation of this signal

Tensor-type couplings are considered in figure \ref{fig:D9-10}. The collider bounds on this coupling are particularly strong, and are again derived from the ATLAS mono-boson search \cite{Aad:2013oja}. They exclude $\Lambda < 2.3$~TeV. Both of the regions favored by the Galactic Center signal, as well as the couplings required for the correct relic density, are comfortably excluded by the LHC result as well as by the direct detection dotted exclusion curves.

Couplings of fermionic DM to gluons are shown in figure \ref{fig:D12-14}. D13, which is not pictured here, leads to purely p-wave annihilations and therefore requires a very low effective scale of about 12 GeV, firmly excluded by the LHC bounds. Both of the remaining operators (D12-D14) give very similar regions of interest for the purposes of explaining the Galactic Center data, and neither conflicts with relic density calculations, but they can potentially be ruled out by the LHC bounds on gluonic interactions of DM, which exclude $\Lambda < 310$~GeV. At such low energies the EFT approach may be broken and therefore I cannot take at face value this limit. The direct detection limits, on the other hand, are quite robust and decisively exclude those operators as viable explanations to the GC excess.

Scalar DM candidates are investigated in figure \ref{fig:scalar}. There are three operators which are not firmly excluded by spin-independent direct detection: C2, C4, and C6. The parameter space for C2 is shown in figure \ref{fig:C2}. The LHC search bound from mono-boson at ATLAS result into the exclusion $\Lambda > 16$~GeV \cite{Aad:2013oja}. Direct detection experiments rule out $\Lambda < 6$~GeV. It is clear that this limit is irrelevant. The C2 ``signal region'' that explains the Galactic Center gamma-ray excess is beyond the reach of LHC searches and is allowed by relic density considerations as well. The operator C4, which is not pictured, gives p-wave suppressed annihilations and requires a suppression scale of about 20 GeV, to be compared with LHC and other bounds at the order of hundreds of GeV. Finally, C6 is considered in figure \ref{fig:C6}. This operator lies above the correct relic density curve, as required for a successful possible explanation of the Galactic Center excess. There has been no LHC search for gluonic couplings of scalar DM candidates, but previous estimates of the most optimistic LHC reach at 14 TeV center of mass energy and 100 fb$^{-1}$ for this operator found suppression scales of order 500 GeV \cite{Goodman:2010ku}, so it is safe to assume that this region is not yet in tension with LHC data.

\section{Conclusions and Outlook}
\label{sec:conc}

We investigated the possibility that the Galactic Center excess in gamma rays could be explained by annihilation of a DM particle whose interactions with the SM are described by higher-dimensional effective operators considering direct detection, collider and abundance arguments. We explored a complete operator basis of all interactions of DM with hadronic matter within the heavy-mediator limit, for both scalar and fermionic (Dirac) DM. We found that there is a set of phenomenologically viable operators capable of producing the right pair-annihilation cross section today to explain the gamma-ray excess. For a scalar DM two options are viable: a scalar-type coupling to quarks, C2, and an operator with gluonic couplings, C6, while for Dirac DM the scalar-type operators D2 and D4 are most promising. Only one vector-type operator, D7, can explain consistently the gamma-ray excess, and only if its couplings are isospin-blind. If the couplings are opposite to up- and down-type quarks the LHC has imposed strict bounds on this operator as well. 

We emphasize that the relevant scales explored in this work might be lower than cut off scale which the EFT approximation is valid, and referred to stringent and robust direct detection constraints when applicable. We believe that these models provide an important benchmark which can be aimed for in DM searches.

Ongoing work on LHC DM searches will continue to improve LHC bounds on effective theories of DM, and ongoing work in understanding the direct detection signatures due to operators which lead to suppressed scattering \cite{Fitzpatrick:2012ib}  will help to shed light on the future detectability of the relevant effective operators with increasingly sensitive direct detection experiments. The effective Hooperon is a viable and testable scenario.

\section*{Acknowledgements}

The authors thank Dan Hooper, Chris Kelso and Eugenio Nobile for helpful discussions. The research of FQ is partly supported by the Brazilian National Counsel for Technological and Scientific Development (CNPq). SP, FQ, and WS are partly supported by the US Department of Energy Award SC0010107. AA is partly supported  Fundacao de Amparao a Pesquisa do Estado de Sao Paulo 2013/22079-8.

\end{document}